# Accurate and scalable multi-element graph neural network force field and molecular dynamics with direct force architecture

Cheol Woo Park[1,2], Mordechai Kornbluth[1], Jonathan Vandermause[3], Chris Wolverton[2], Boris Kozinsky[3,1*], Jonathan P. Mailoa[1*]

1) Robert Bosch Research and Technology Center, Cambridge, MA 02139, USA
2) Northwestern University, Evanston, IL 60208, USA
3) Harvard School of Engineering and Applied Sciences, Cambridge, MA 02138, USA

* corresponding authors: jpmailoa@alum.mit.edu, bkoz@seas.harvard.edu

## Abstract

Recently, machine learning (ML) has been used to address the computational cost that has been limiting *ab initio* molecular dynamics (AIMD). Here, we present GNNFF, a graph neural network framework to directly predict atomic forces from automatically extracted features of the local atomic environment that are translationally-invariant, but rotationally-covariant to the coordinate of the atoms. We demonstrate that GNNFF not only achieves high performance in terms of force prediction accuracy and computational speed on various materials systems, but also accurately predicts the forces of a large MD system after being trained on forces obtained from a smaller system. Finally, we use our framework to perform an MD simulation of $Li_7P_3S_{11}$, a superionic conductor, and show that resulting Li diffusion coefficient is within 14% of that obtained directly from AIMD. The high performance exhibited by GNNFF can be easily generalized to study atomistic level dynamics of other material systems.

## I. Introduction

In the past few decades, molecular dynamics (MD) has been extensively used to study and understand a wide range of chemical/physical phenomena at the atomistic level. The movement of the atoms in an MD simulation is dictated by the force fields obtained as gradients of the potential energy surface (PES) of the system. A common way of obtaining the PES is to explicitly calculate the electronic structure of the simulated system through *ab initio* approaches; density functional theory (DFT) is one of the methods in this category balancing speed and accuracy [1-3]. While *ab initio* MD (AIMD) provides highly accurate dynamics for many different atomic systems, the high computational cost of calculating the electronic structure derived forces places a hard limit on the system size (number of atoms) and duration (timesteps) of the simulation.

An alternative to AIMD is classical MD where the PES of an atomic system is constructed using interatomic potentials that define the interactions between atoms under specific bonding environments. Because it bypasses the need to calculate the electronic structure of the simulated system, classical MD has a significantly higher computational speed than AIMD and can be used to simulate larger systems for longer durations. Over the years, many different interatomic potentials have been developed to describe different bonding environments [4-16] such as the Buckingham potential [17] that describes ionic systems or the embedded-atom method [4] that describes metallic systems. However, each potential is generally constrained to the one or several bond types that it was designed for, making these potentials unsuitable for simulating systems that have diverse and dynamic bonding environments often occurring in many important applications such as batteries and fuel cells [18-20]. While interatomic potentials like the reactive force field (ReaxFF) [13] are actively being developed to simulate such complicated bonding environments, these methods, in their current states, are still limited by accuracy [21, 22].

To address the limitations of both classical and *ab initio* MD, there has been an increasing interest in using machine learning (ML) to generate *ab initio* quality MD with the computational efficiency comparable to that of classical MD [23-31]. The majority of the current state-of-the-art ML models use a set of features, often referred to as "atomic fingerprints," to represent the local environments of atoms that constitute the system of interest [23, 32-38]. These features are generally rotationally- and translationally-invariant, i.e. they remain constant under arbitrary rotational or translational transformations of the coordinate space. The atomic fingerprints are used to predict the PES of the system prior to taking its spatial derivative to obtain the force fields. The fingerprints are derived from manually designed functions that take as input the position of the center atom for which the fingerprint is being computed and the positions of neighboring atoms that are within a certain proximity of the center atom [23, 33].

Although the computational efficiencies of these ML models are better than *ab initio* methods, the cost of using these models in practice is still significantly high due to the computational bottlenecks that come from 1) deriving the atomic fingerprints needed for PES predictions and 2) calculating the derivatives of the PES to obtain forces [33, 39]. More recently developed models address one or the other of these computational bottlenecks, but not both. In one approach, models bypass the need to design and compute the atomic fingerprints by utilizing deeper neural network architectures, similar to convolution or graph neural networks [40, 41], to automatically extract structural information of a material system [28-31, 38, 42-45]. For example, SchNet [28, 30] uses a series of continuous-filter convolutional layers to extract features from an arbitrary molecular geometry that can then be used to predict the PES of the molecule. While these models have shown excellent accuracies of atomic forces in single molecule systems, these models still rely on taking the derivatives of the PES to predict the atomic forces. In a different approach, the Direct Covariant Forces (DCF) model [39] proposes an ML framework that predicts atomic forces directly using atomic fingerprints without having to take the derivatives of the PES. However, because the model relies on human-engineered atomic fingerprints, its prediction accuracy is limited.

In this study, we present a new graph neural network force field (GNNFF) framework that bypasses both computational bottlenecks by predicting atomic forces directly using automatically extracted structural features that are translationally-invariant, but rotationally-covariant to the coordinate space of the atomic positions. We first demonstrate the accuracy and speed of GNNFF in calculating the atomic forces of various material systems against state-of-the-art models and benchmarks developed for each of those systems. For the first benchmark system of organic molecules found in the ISO17 database [28, 29, 46], we compare GNNFF against the SchNet architecture. GNNFF outperformed SchNet by 16% in force prediction accuracy and by a factor of 1.6x in prediction speed. For the second benchmark system consisting of two multi-element solid-state systems (amorphous $Li_4P_2O_7$ and $Al_2O_3$ in contact with high-concentration HF gas at the surface), a class of materials often neglected in ML tests, we benchmarked GNNFF against the DCF architecture [39]. Depending on the element-type and the system, GNNFF outperformed DCF by up to 30% on force prediction accuracy (up to 6.2x higher accuracy for rare chemical reaction events particularly difficult for DCF) and 4.5x on prediction speed.

We then further characterize the practical aspects of GNNFF's performance which are based on 1) scalability and 2) the physical/chemical accuracy of the GNNFF generated trajectories. Scalability measures how accurately the ML model can predict the forces of a system after being trained on forces obtained from a smaller system. For this assessment, we generated two separate AIMD trajectories of $Li_{7-x}P_3S_{11}$, a superionic conducting material. One uses a 1x2x1 supercell (Small) and the other a 1x2x2 supercell (Large), each with a single Li-ion vacancy compensated by a background charge. When trained solely on the forces taken from the "Small" supercell, we show that GNNFF is able to predict the forces of the "Large" system as accurately as it can predict the forces of the "Small" system where the differences in accuracies in predicting the forces were within 3%. To illustrate the physical/chemical accuracy of the GNNFF generated trajectories, we measured the Li-ion diffusion coefficient from an ML-driven simulation of $Li_{7-x}P_3S_{11}$ performed under the same conditions as the AIMD "Small" simulation using the previously trained GNNFF. The diffusivity measured from GNNFF-generated trajectory was within 14% relative to that measured directly from the AIMD "Small" trajectory. The versatility and excellent performance exhibited by GNNFF strongly suggest that it can be used to effectively accelerate many studies of atomistic level dynamics through MD.

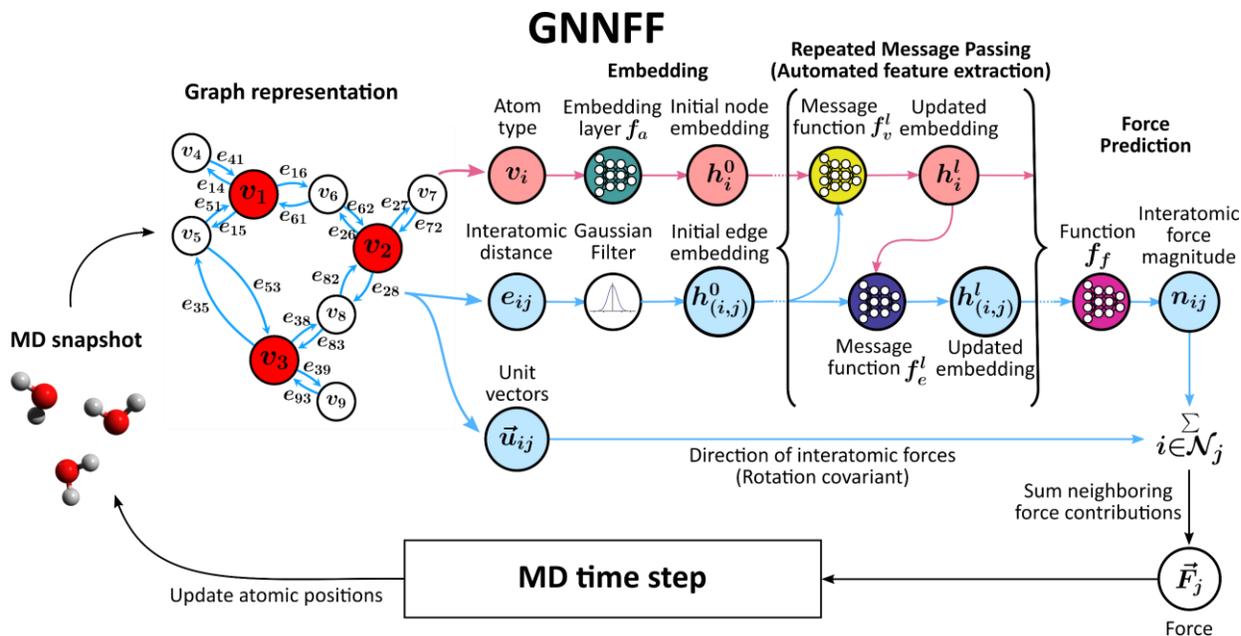

**Figure 1. Illustration of the GNNFF architecture.** Each atomic configuration is represented as a directed graph that is then used as the input for GNNFF. Atom $i$ is represented as node $v_i$ and the influence atom $i$ has on atom $j$ is represented by a directed edge $e_{ij}$. The direction of this influence is represented by unit vector $\vec{u}_{ij}$. Nodes and edges are embedded with latent vectors in the embedding stage. Initially, the node and edge embeddings respectively contain the atom type and interatomic distance information. The embeddings are then iteratively updated during the message passing stage. The final updated edge embeddings are used for predicting the interatomic force magnitudes. Force field of center atom $j$ is calculated by summing the force contributions of neighboring atoms $i \in \mathcal{N}_j$ that are calculated by multiplying the force magnitude and the respective unit vector. The predicted forces are finally used for updating the atomic positions in MD.

## II. The Graph Neural Network Force Field (GNNFF) model

The general architecture of GNNFF is illustrated in Figure 1, and is inspired by previous work on crystal graph convolutional neural networks [31, 45]. First, the atomic structure of each MD snapshot is represented as a directed graph $\mathcal{G} = (\mathcal{V}, \mathcal{E}, \mathcal{U})$, with node $v_i \in \mathcal{V}$ representing the constituent atom $i$, directed edge $e_{ij} \in \mathcal{E}$ representing the influence atom $i$ has on neighboring atom $j$, and $\vec{u}_{ij} \in \mathcal{U}$ representing the unit vector pointing from atom $i$ to atom $j$. Here, we emphasize that the edges are directed where $e_{ij} \neq e_{ji}$ implying that the influence that atom $i$ has on atom $j$ is not the same as the influence that atom $j$ has on atom $i$. Each node is connected to its $N$ closest neighbors where $N$ is a predefined number. The graph representation is used as input for GNNFF which is composed of the embedding, the message-passing, and the force vector calculation stages.

In the embedding stage, node $v_i$ and edge $e_{ij}$ are embedded with latent vectors to represent the hidden physical/chemical states of the atom and bond respectively, where we define embedding to be the mapping of a discrete object to a vector of real numbers. Initial node embedding $h_i^0$ is the output of a neural network where the input is a one-hot (categorical) vector representing the element-type of atom $i$. The purpose of using one-hot vector representations is purely to differentiate one element-type from another. For example, in a system that consists only of water, hydrogen and oxygen can be respectively represented as one-hot vectors [1, 0] and [0, 1]. The initial edge embedding $h_{(i,j)}^0$ is given by expanding the distance between atoms $i$ and $j$ in a Gaussian basis, a process we refer to as Gaussian filtering. Mathematically this can be expressed as:

$$f_a: \text{One-Hot}(Type(\text{atom } i)) \rightarrow h_i^0$$

$$G: d(i,j) \rightarrow h^0_{(i,j)}$$

where $f_a$ denotes the neural network that maps the one-hot encoding of the element-type of atom $i$ to $h^0_i$, $G$ denotes the Gaussian filtering operation, and $d(i,j)$ denotes the distance between atoms $i$ and $j$.

The message-passing stage of GNNFF consists of multiple layers where in each layer, "local messages" are passed between neighboring nodes and edges. We define the local message passing that occurs between the nodes and edges as follows:

$$\text{Edge to node: } f^l_v: \{h^l_i, h^l_{(i,j)} | j \in \mathcal{N}_i\} \rightarrow h^{l+1}_i$$

$$\text{Node to edge: } f^l_e: \{h^{l+1}_i, h^{l+1}_j, h^l_{(i,j)}\} \rightarrow h^{l+1}_{(i,j)}$$

$h^l_i$ and $h^l_{(i,j)}$ are the embeddings of node $v_i$ and edge $e_{ij}$ in the message-passing layer $l$, respectively. Initial conditions for $h^l_i$ and $h^l_{(i,j)}$ are given by $h^0_i$ and $h^0_{(i,j)}$ which were determined in the embedding stage. $\mathcal{N}_i$ denotes the set of indices of neighboring nodes connected to node $v_i$. Message functions $f^l_v$ and $f^l_e$ are node- and edge- specific neural networks that update $h^l_i$ and $h^l_{(i,j)}$ to $h^{l+1}_i$ and $h^{l+1}_{(i,j)}$ respectively. These functions are defined such that after each update (message-passing), embedding $h^l_i$ better represents the local atomic environment of atom $i$ and embedding $h^l_{(i,j)}$ better represents the interatomic interactions between atoms $i$ and $j$. The designs for these functions are flexible and can be altered to emphasize the different interatomic interactions that occur within MD simulations. In the current implementation of GNNFF, we defined the functions to be the following:

$$f_v^l: h_i^{l+1} = g\left[h_i^l + \sum_{j\in\mathcal{N}_i} \sigma\left((h_i^l \oplus h_{(i,j)}^l)W_1^l + b_1^l\right) \odot g\left((h_i^l \oplus h_{(i,j)}^l)W_2^l + b_2^l\right)\right] \quad (1)$$

$$f_e^l: h_{(i,j)}^{l+1} = g\left[h_{(i,j)}^l + \sigma\left((h_i^l \odot h_j^l)W_3^l + b_3^l\right) \odot g\left((h_i^l \odot h_j^l)W_4^l + b_4^l\right) + \right.$$
$$\left. \sum_{k\in\mathcal{N}_j} \sigma\left((h_i^l \oplus h_j^l \oplus h_k^l \oplus h_{(i,j)}^l \oplus h_{(k,j)}^l)W_5^l + b_5^l\right) \odot g\left((h_i^l \oplus h_j^l \oplus h_k^l \oplus h_{(i,j)}^l \oplus h_{(k,j)}^l)W_6^l + b_6^l\right)\right]$$
$$(2)$$

where $g$ and $\sigma$ respectively denote a hyperbolic tangent function and a sigmoid function that are applied element-wise to each entry of the vector. $\odot$ denotes an element-wise multiplication operator, and $\oplus$ denotes a concatenation operator. $W^l$ and $b^l$ represent the weights and biases of the neural network hidden layers that compose the message-passing layer $l$. In Eq 1., $h_i^l \oplus h_{(i,j)}^l$ represents how the interatomic dynamic between atoms $i$ and $j$ impacts the local environment of atom $i$. In Eq 2., $h_i^l \odot h_j^l$ and $h_i^l \oplus h_j^l \oplus h_k^l \oplus h_{(i,j)}^l \oplus h_{(k,j)}^l$ respectively represent the 2-body correlation between atoms $i$ and $j$ and 3-body correlation between atoms $i$, $j$, and $k$. We note that the 2-body and 3-body correlation terms use different operators. These operators were chosen based on trial-and-error where we found that using the $\odot$ operator for the 2-body correlation term and $\oplus$ operator for the 3-body correlation term yielded the best ML performance. The many-body correlations beyond the 2-body and 3-body contributions are implicitly captured by repeatedly passing messages between the nodes and edges. However, this can also result in the loss of information as the message travels further away from its point of origin. To minimize this loss, Eq 1. and Eq 2. use an activation function $g$ to normalize information after each message-passing iteration and an activation function $\sigma$ to act as a gate that allows only the most relevant information to pass through [47].

The use of latent vectors that are iteratively updated instead of human-engineered atomic fingerprints to represent the chemical/physical state of the local atomic environment in itself is not a new practice [28, 30, 31, 45]. The novelty of GNNFF comes from how it uses the updated edge embeddings to calculate force vectors that are rotationally-covariant to the coordinate space in the final stage. For a GNNFF with $L$ message-passing layers, the final updated state of edge $e_{ij}$, represented by embedding $h^L_{(i,j)}$, is used to calculate $n_{ij}$, a scalar quantity that denotes the magnitude of the force contribution that atom $i$ is exerting on atom $j$. The force contribution of atom $i$ onto $j$ is given by $n_{ij}\,\vec{u}_{ij}$ and the total force prediction on atom $j$ is simply the vector sum of the force contributions of all neighboring atoms. Mathematically the individual force contribution and the total force prediction can respectively be written as:

$$f_f: h^L_{(i,j)} \to n_{ij}$$

$$\vec{F}_j = \sum_{i \in \mathcal{N}_j} n_{ij}\,\vec{u}_{ij}$$

where the function $f_f$ is a neural network in the final stage that maps $h^L_{(i,j)}$ to $n_{ij}$ and $\vec{F}_j$ is the force prediction on atom $j$. Because $\vec{F}_j$ is a linear function of $\vec{u}_{ij} \in \mathcal{U}$, which is inherently translationally-invariant and rotationally-covariant to the input coordinates of the atoms, the GNNFF-predicted force fields are translationally-invariant and rotationally-covariant to the input atomic coordinates environment $\{\vec{R}_t\}$ of the MD snapshots as well. Also, since we only consider the force contributions of a fixed number of neighboring atoms, the computational cost of using GNNFF to predict new forces scales linearly with the size of the system. The weights and biases in $f_a$, $f_v^l$, $f_e^l$, and $f_f$ are shared across all the atoms.

## III. Results

## A. Evaluation of force predictive accuracy of GNNFF and comparison to existing models

In this section, we characterize the performance of GNNFF in predicting atomic forces through two different illustrations. First, we evaluate GNNFF performance in predicting the forces of simple organic molecules in reference to SchNet which provides one of the best published benchmarks available for predicting forces of single-molecule MD [28, 30]. Subsequently, GNNFF is evaluated on complex solid-state systems in reference to DCF, which has been trained and tested for predicting forces of complicated multi-element solid-state systems [39].

### 1. Using GNNFF and SchNet to predict atomic forces of simple organic molecules

To compare the GNNFF performance in predicting the forces of organic molecules to that of SchNet, GNNFF was trained and tested in the same way SchNet was evaluated in Ref [28]. The organic molecule force data used for evaluation were taken from the ISO17 database [28, 29, 46], a collection of MD trajectories simulated for 129 organic isomers, all with the composition of $C_7O_2H_{10}$, but with distinct structures. The trajectory for each isomer consists of 5000 snapshots, resulting in a total of 645,000 unique snapshots. The snapshots are divided into three non-overlapping sets: one training set and two test sets. The training set consists of 80% of the snapshots randomly selected from the MD trajectories of 80% of the 129 isomers. The first test set, referred to as the "test within" set, consists of the remaining 20% of the snapshots obtained from the same molecules used in the training set. The second and more challenging test set consists of all the snapshots taken from the other 20% of the 129 molecules that are not included in the training set and is referred to as the "test other" set. The purpose of the first test set is to evaluate an ML model's ability to interpolate the force fields of unknown geometries for known molecules, i.e. molecules that are present in the training data, while the second test set evaluates the ability of GNNFF to generalize force fields for unknown molecules, i.e. molecules that were never encountered in training. Predictive accuracies are evaluated based on the mean absolute error (MAE) of the Cartesian force components.

**Table 1.** Comparison of GNNFF and SchNet Cartesian force components predictive accuracy of the ISO17 database. The mean absolute value of all atomic forces, $\langle|\vec{F}_{DFT}|\rangle$, for each test set is shown for reference.

| Test set | $\langle|\vec{F}_{DFT}|\rangle$ (eV/Å) | MAE (eV/Å) | |
|---|---|---|---|
| | | SchNet [28] | GNNFF |
| test within | 1.698 | 0.043 | **0.036** |
| test other | 1.664 | 0.095 | **0.088** |

The test results are summarized in Table 1. For "test within", MAE of GNNFF was 0.036 eV/Å which is 16% lower than that of SchNet (0.043 eV/Å) indicating that GNNFF is better at capturing chemical/structural features of a molecular conformation necessary for interpolating the force fields of unknown conformations for known molecules, i.e. molecules that are included in the training set. For "test other", GNNFF MAE of 0.088 eV/Å was 7% lower compared to that of SchNet (0.095 eV/Å) implying that GNNFF extracted chemical/structural features are more generalized and therefore, better suited for inferring the force fields of new unknown molecules that the ML model never encountered in training.

We also evaluated the computational efficiency of SchNet and GNNFF in predicting atomic forces of the "test within" and "test other" data sets. Evaluations were performed on a workstation equipped with an Nvidia GTX 1080 GPU and an Intel i7-8700K 6-core 3.70GHz CPU processor. The efficiency of SchNet was evaluated using the SchNetPack code that is available through github.com/atomistic-machine-learning/schnetpack. The time for SchNet to evaluate the forces of all the MD snapshots in the "test within" and "test other" sets respectively was $9.8 \times 10^{-4}$ and $1.1 \times 10^{-3}$ s/atom/snapshot. For GNNFF, these times were respectively $6.0 \times 10^{-4}$ and $6.1 \times 10^{-4}$ s/atom/snapshot, indicating that GNNFF is approximately 1.6x faster, most likely because GNNFF, unlike SchNet, does not need to predict the energies of the molecules prior to calculating the forces. These results demonstrate that GNNFF can be effectively used for accelerating the chemical exploration of organic molecules with performance on par with current state-of-the-art models.

**2. Using GNNFF and DCF to predict atomic forces of complex solid-state systems: $Li_4P_2O_7$ and $Al_2O_3$-HF**

Here, we benchmark the GNNFF performance on the AIMD trajectories of two different solid-state systems. The first system is $Li_4P_2O_7$ at a temperature T=3000K, well beyond its melting temperature. Temperature was regulated with a Nosé-Hoover thermostat and MD was performed with a timestep of 2 fs for 50 ps yielding a total of ~25,000 snapshots. This simulation represents part of the rapid annealing process that turns crystalline $Li_4P_2O_7$ into an amorphous state. Using this system, we evaluate GNNFF's ability to learn and recognize the subtle differences between the different phases of the oxide that are present in the simulation. The second system involves $Al_2O_3$ in contact with HF molecules at T=1000K where chemical reactions occur between the HF acids and the Al atoms. Temperature was regulated with a Nosé thermostat and a timestep of 0.5 fs was used with a total duration of ~7 ps yielding ~13,000 snapshots. This system is used for evaluating GNNFF's performance in capturing dynamics that drive chemical reactions. The $Li_4P_2O_7$ and $Al_2O_3$-HF systems consist of 208 and 228 atoms respectively, both significantly larger than the isomers of ISO17. Detailed explanations on the generation of these simulations can be found in Ref [39]. GNNFF was trained and tested separately for each MD trajectory where the snapshots were shuffled and divided into 80% and 20% of the total data to form the training set and test set. For the $Li_4P_2O_7$ system, the training data consisted of ~20,000 snapshots and the testing data consisted of ~5,000 snapshots. For the $Al_2O_3$-HF system, there were ~11,000 training and ~3,000 testing snapshots. Similar to the DCF evaluation criteria [39], the GNNFF performance was quantified separately for each element-type $e$ by the ratio of the vector mean absolute error (vMAE) between the GNNFF-predicted forces ($\vec{F}_{GNNFF,e}$) and DFT-calculated forces ($\vec{F}_{DFT,e}$), to the mean absolute value (MAV) of the DFT-calculated forces. Specifically, prediction accuracy is given by

$$\frac{vMAE_e}{MAV_e} = \frac{\langle |\vec{F}_{DFT,e} - \vec{F}_{GNNFF,e}| \rangle}{\langle |\vec{F}_{DFT,e}| \rangle},$$

a metric that is comparable across temperatures and elements where the brackets indicate an averaging operation. The numerator averages the absolute differences between the DFT-calculated and GNNFF-predicted forces for element-type $e$ while the denominator averages the magnitude of the DFT-calculated forces of $e$. Lower $\frac{vMAE_e}{MAV_e}$ indicates higher predictive accuracy. We note that this is different from the MAE of the Cartesian force components used previously to compare GNNFF and SchNet.

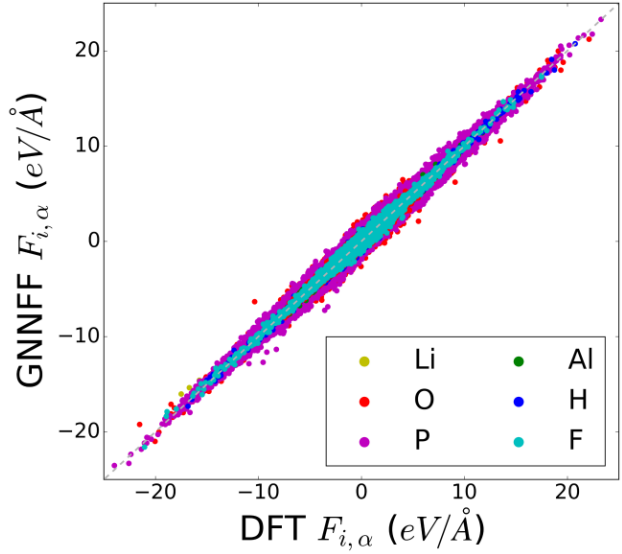

**Figure 2. Performance of GNNFF in predicting forces of solid-state systems.** GNNFF-predicted vs DFT-calculated force correlation plots of elements from the Li$_4$P$_2$O$_7$ and Al$_2$O$_3$-HF MD systems. $F_{i,\alpha}$ ($\alpha \in \{x, y, z\}$) represents the Cartesian force component of atom $i$.

As shown in Figure 2, the correlations between the ML-predicted and DFT-calculated forces are clearly evident for all elements in both systems, implying the high general performance of GNNFF in predicting forces. The predictive accuracy of GNNFF for each element is summarized in Table 2 and compared to that of DCF. The improvement of GNNFF over DCF measured in vMAE/MAV % values ranges from 8% for the lighter elements such as H, O, and F to 30% for the heavier elements such as P and Al.

**Table 1.** Comparing the accuracies of GNNFF and DCF in predicting the atomic forces of solid-state systems, $Li_4P_2O_7$ and $Al_2O_3$-HF. Accuracy is measured separately for each element in terms of vMAE/MAV, where lower values indicate higher accuracies. For both systems, GNNFF accuracy for each element is higher than that of DCF.

| System | Atom | vMAE/MAV (%) | |
|---|---|---|---|
| | | DCF | GNNFF |
| $Li_4P_2O_7$ | Li | 37 | 13 |
| | O | 25 | 10 |
| | P | 40 | 10 |
| $Al_2O_3$-HF | Al | 33 | 14 |
| | F | 22 | 14 |
| | H | 22 | 14 |
| | O | 35 | 15 |

The time for GNNFF to evaluate the test sets for the $Li_4P_2O_7$ and $Al_2O_3$-HF systems were both $1.8 \times 10^{-3}$ s/atom/snapshot on a single core of a high-performance cluster mounted with 2.10GHz Intel Xeon Gold 6230 processor. Using the same hardware setup, DCF took $7.8 \times 10^{-3}$ and $8.8 \times 10^{-3}$ s/atom/snapshot to evaluate the $Li_4P_2O_7$ and $Al_2O_3$-HF test sets respectively, showing that GNNFF is on average 4.5x faster than DCF in predicting forces. These results again demonstrate GNNFF's competitive edge against existing state-of-the-art models.

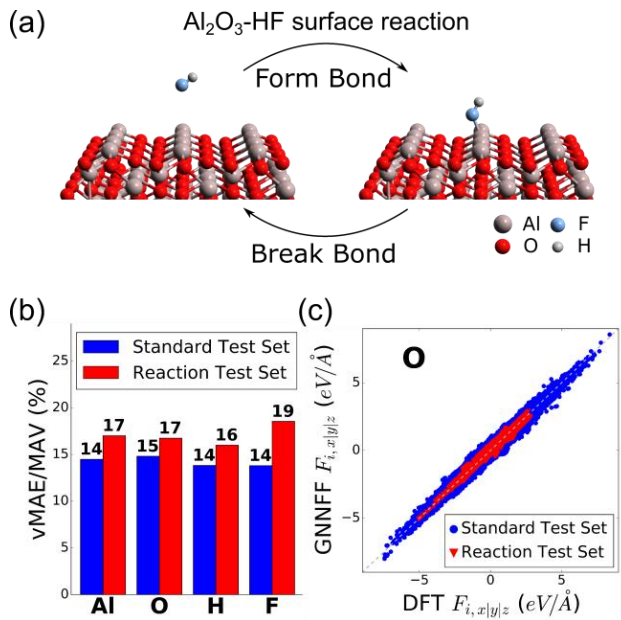

**Figure 3. Predicting the force fields of atoms involved in surface reaction between Al$_2$O$_3$ and HF.** (a) Illustration of the Al$_2$O$_3$-HF surface reaction (b) Comparison of force predictive accuracies between standard and reaction test sets for all elements present in system. (c) GNNFF-predicted vs DFT-calculated force correlation plots of oxygen for both standard and reaction test sets.

Using the Al$_2$O$_3$-HF system, we further investigated the GNNFF's performance in predicting the atomic force fields involved in rare events as was done in Ref [39]. The trajectory contains chemical reactions where a HF molecule associates and dissociates from an Al atom at the surface of Al$_2$O$_3$ as shown in Figure 3(a). The snapshots that correspond to the events in which the HF molecules react with the Al atoms, or simply the "reaction snapshots", occur on average once every 8 time steps and are extracted from the trajectory according to a Hidden Markov Model (HMM) as explained in the Methods section.

While the sparsity of data makes it especially difficult to train an ML model to correctly predict the atomic forces of reacting snapshots, it is nonetheless important to quantify an ML model's performance in this regard as these rare occurrences are generally the phenomena we would be interested in studying with ML-driven MD. For our assessment, we trained another GNNFF model from scratch using only the nonreacting snapshots and tested it using the reacting snapshots (reaction test set). There were ~1,600 reacting snapshots leaving ~12,400 snapshots for training.

We compared the force prediction accuracy of GNNFF on the reaction test set to our previous evaluation where snapshots of the $Al_2O_3$-HF were randomly shuffled prior to being divided into training and testing data (standard testing set) as reported in Table 2. Figure 3(b) shows that the vMAE/MAV % values for each element-type measured on the reaction and standard test set differ by 5% or less. According to the HMM that is used for extracting the reaction snapshots, oxygen is occasionally recognized as part of the alumina surface and HF reaction when the HF acid interacts with a unit of Al-O rather than the individual Al atom. However, the reaction snapshots corresponding to these reaction events involving oxygen comprise less than ~0.1% of the total trajectory, making force predictions of oxygen in the reaction test set a challenge [39]. In Figure 3(c), the correlation between the GNNFF-predicted and DFT-calculated forces of the reaction test set is as clear as that of the standard test set. The GNNFF vMAE/MAV % value for oxygen in the reaction test set was 17%. The corresponding value reported for DCF is 6.2x higher at 105%. We hypothesize that this significant improvement happens because GNNFF is a fully rotationally-covariant force prediction algorithm that does not require hand-crafted descriptors, while DCF is only trained to be rotationally-covariant using data augmentation and relies on manually designed descriptors.

## B. Assessing the practical effectiveness of GNNFF

The expected effectiveness of GNNFF in practice can be further tested based on two additional criteria. The first criterion is scalability: how does the GNNFF perform when we use it to predict the force fields of an MD system that is larger than the system used for generating the training data? *Ab initio* methods are limited by computational cost in simulating large materials systems with a lot of atoms. This can be problematic when studying physical phenomena such as nucleation, dislocation, and grain boundary formations which require sufficiently large MD systems in order to be observed. Having an ML model that can consistently provide accurate force field predictions regardless of the MD system size would greatly expand our capability to explore diverse physical phenomena at the atomic level. The second criterion is how physically/chemically accurate the GNNFF generated simulations are in reference to AIMD [48, 49]. For example, if we are interested in studying the ionic conducting properties of a material using ML, it is important that the ionic conductivity measured from the ML simulated trajectory to be close to that measured from AIMD.

Our system of choice to test GNNFF for these criteria is $Li_7P_3S_{11}$, a superionic conducting material. Recent study has shown that the conductivity of $Li_7P_3S_{11}$ at 300K predicted by AIMD is 57 mS/cm which is about 5x greater than the experimentally measured 11.6 mS/cm [19]. The lower experimentally measured conductivity is speculated to be due to grain boundaries which deter ion transport. Using AIMD to accurately determine the effect that grain boundaries (GB) have on ion transport is problematic, as it would require significantly larger systems than bulk (GB free) simulations. While a detailed investigation of ion dynamics in grain boundaries using ML is beyond the scope of this study, $Li_7P_3S_{11}$ is a good example where having an ML model that satisfies the aforementioned scaling and accuracy criteria could greatly help in studying its physical/chemical properties.

Data used for this assessment was obtained from two separate AIMD trajectories of $Li_{7-x}P_3S_{11}$. These trajectories differ in system size, where one uses a 1x2x1 supercell ("Small") and the other uses a 1x2x2 supercell ("Large"). Both systems contain a single $Li^+$ vacancy to accelerate ion diffusion resulting in 83 atoms (x=0.25) in the "Small" system and 167 atoms (x=0.125) in the "Large" system. Simulations were performed at 520K using a Nosé-Hoover thermostat. The "Small" and "Large" trajectories respectively consist of ~25,000 and ~7,500 snapshots. More details of how these simulations were generated can be found in the Methods section.

## 1. Evaluating GNNFF scalability

To illustrate the scalability of GNNFF, we first trained it on ~20,000 snapshots randomly selected from the "Small" trajectory. We then compared GNNFF accuracy in predicting the forces of all ~7,500 snapshots provided by the "Large" trajectory and the ~5,000 snapshots in the "Small" trajectory that were not included in the training data. Accuracy was again measured separately for each element-type in vMAE/MAV.

As shown in Figure 4(b), the differences in vMAE/MAV % values for the "Small" and "Large" systems were within 3% for all elements, demonstrating the consistency of the GNNFF force prediction accuracy on the "Large" trajectory even after being trained on the "Small" trajectory forces. In Figure 4(c), the GNNFF-predicted vs DFT-calculated force correlation plots of the element S for the "Large" and "Small" test sets are overlaid together. While the force correlation of the "Large" test set is as good as that of the "Small" test set, some of the ML-predicted values significantly deviate from the DFT-calculated values as indicated by the blue arrows. Although these deviations were only observed for the force predictions of S, they could cause concern as they might lead to unphysical atomic configurations when using GNNFF to perform MD.

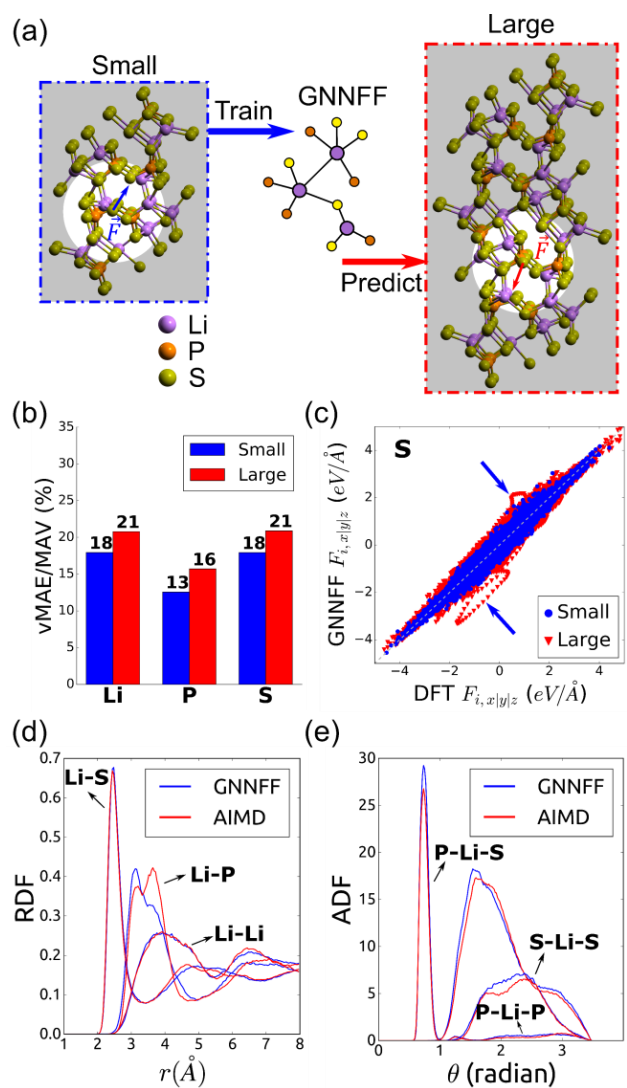

**Figure 4. Evaluating GNNFF scalability on AIMD trajectories of $Li_{7-x}P_3S_{11}$ with differing x values.** (a) GNNFF is trained on "Small" trajectory (x=0.25) forces and tested on "Large" trajectory (x=0.125) forces (b) Comparison of GNNFF performance when evaluated on "Small" trajectory forces not used for training vs "Large" trajectory forces. (c) GNNFF-predicted vs DFT-calculated force correlation plots of S for both "Small" and "Large" systems. Blue arrows indicate significant deviations of ML-predicted values from DFT-calculated values. (d) (e) Comparison of the (d) radial distribution functions of different element pairs involving Li and (e) angular distribution functions of different

element triplets involving Li from GNNFF and AIMD generated "Large" trajectories. The second element is the central atom of the triplet.

To address this concern, we performed an NVT MD simulation for the "Large" system size, but using atomic forces (and hence, evolution of the trajectory) calculated by the GNNFF trained on the "Small" trajectory forces. We then compared atomic configurations of the GNNFF-simulated trajectory to the *ab initio* generated "Large" trajectory. For a fair comparison, GNNFF-generated simulation was performed under the same conditions as its *ab initio* counterpart in terms of temperature, thermostat, timestep, simulation duration, and initial structure. Throughout the ML-driven simulation, no breaking of the P-S bonds was observed. The radial distribution functions (RDFs) and angular distribution functions (ADFs) of the GNNFF and AIMD trajectories are highly consistent with one another as shown respectively in Figures 4(d) and (e). This demonstrates that the average atomic configurations generated by GNNFF MD are physically reasonable despite the deviations in the force predictions observed in Figure 4(c).

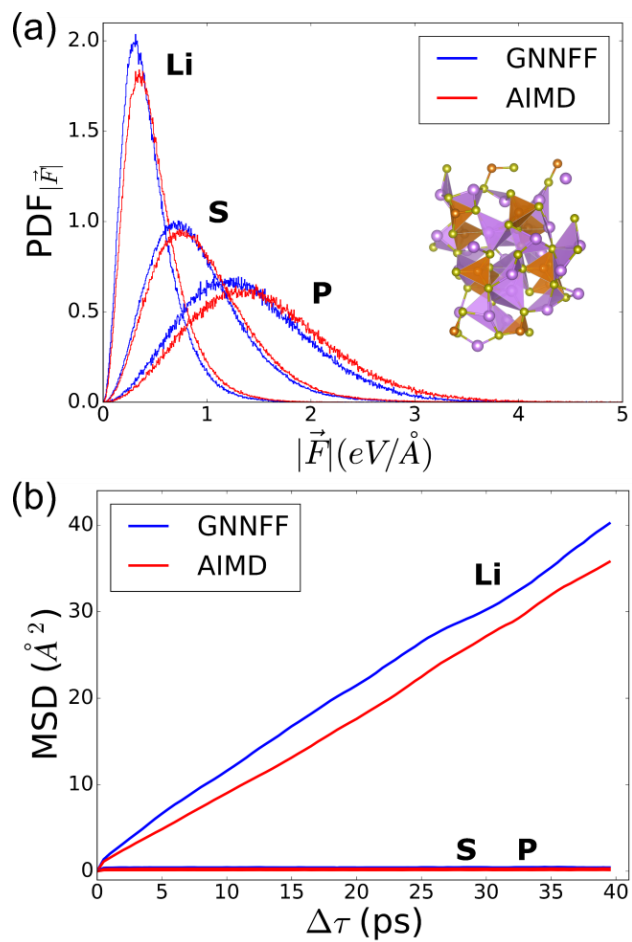

**Figure 5. Evaluating physical/chemical accuracy of GNNFF-generated trajectory of Li$_{7-x}$P$_3$S$_{11}$ in comparison to AIMD equivalent.** (a) Probability distribution functions of force magnitudes for each element. Inset figure shows the structure of Li$_{7-x}$P$_3$S$_{11}$ obtained from the GNNFF-MD run after 50 ps. (b) Mean squared displacement of Li, S, and P.

**2. Evaluating physical/chemical accuracy of GNNFF-generated trajectories**

For this evaluation, we calculate the diffusivity of Li-ions in Li$_{7-x}$P$_3$S$_{11}$ (x=0.25) with an MD simulation using GNNFF-predicted forces and compare it with that obtained directly from AIMD. The *ab initio* value for the Li diffusivity was obtained from the AIMD "Small" trajectory that was simulated for 50 ps. For an apples-to-apples comparison, we performed an MD simulation equivalent of the "Small" trajectory using the previously trained GNNFF to determine the ML predicted Li diffusivity. Again, no breaking of the P-S bond was observed throughout the GNNFF-driven simulation. In Figure 5(a), we compare the element-specific probability distribution function (PDF) of the force magnitudes in the GNNFF MD run and AIMD run. While there is an overall tendency of GNNFF to underestimate the force magnitudes compared to DFT, we generally see good agreement between the GNNFF and AIMD PDFs for all element-types. In Figure 5(b), we compute and plot the mean square displacement (MSD) of Li-ions as a function of time lag ($\Delta\tau$). The diffusivity of $Li_{7-x}P_3S_{11}$ (x=0.25) at 520K determined from the MSD slope was $1.7 \times 10^{-5}$ $cm^2$/s for GNNFF which is in good agreement with the AIMD value of $1.5 \times 10^{-5}$ $cm^2$/s. This demonstrates that the accuracy of the GNNFF-driven MD simulations is on par with that of the simulations generated through *ab initio* methods. With methods like GNNFF that are more scalable than AIMD, direct simulations of ion transport across defects and grain boundaries are more feasible and will be the subject of future work.

## Conclusions

GNNFF provides a novel versatile framework to directly predict atomic forces from automatically extracted features of the local atomic environment that are rotationally-covariant to the coordinate space. This enables GNNFF to bypass both computing atomic fingerprints and calculating the derivatives of the PES which are often the computational bottlenecks of existing ML force field models.

In this work, we demonstrated the competitive edge of GNNFF in terms of accuracy and speed with respect to existing models such as SchNet and DCF in predicting the forces of various systems ranging from simple organic molecules and complex solid-state systems. We further showed that GNNFF has good scalability and can accurately predict the forces of large systems even after being trained on forces from a smaller system. Finally, we validated the physical/chemical accuracy of GNNFF by showing that the MD simulation of $Li_{7-x}P_3S_{11}$ driven by GNNFF-calculated forces can quantitatively capture the dynamics of Li-ion diffusion. We envision that GNNFF will significantly accelerate many different atomistic studies with its scalability, flexibility, and excellent force prediction accuracy.

## Methods

**Extracting reaction snapshots from the $Al_2O_3$-HF MD trajectory**

The reaction snapshots in the $Al_2O_3$-HF MD trajectory that correspond to the HF acid reacting at the surface of $Al_2O_3$ were extracted using an approach that combines molecular graph analysis and a Hidden Markov Model (HMM) [50, 51]. In this approach, snapshots of an MD trajectory are represented as graphs where each atom is connected to all neighboring atoms within a radial cutoff distance that depends on the covalent radii of the atoms. Here, we note that a connection between two atoms does not necessarily mean that there exists a bond between them. A connection is classified as a bond only after the connection consistently exists throughout a continuous time span. The level of consistency for which the connection must exist to be classified as a bond is determined by the HMM. This approach ensures that we do not mistake a cluster of atoms to be of the same molecule when they just happened to be close to each other due to thermal vibration or molecular collision. More details of this approach can be found in Ref [39].

**Generating $Li_{7-x}P_3S_{11}$ *Ab Initio* Data**

The AIMD simulations of $Li_{7-x}P_3S_{11}$ were performed using the Vienna *Ab Initio* Simulation Package (VASP) [2, 3] with the project augmented wave (PAW) method [52] and Perdew-Burke-Ernzerhof (PBE) generalized gradient approximation [53]. Two separate MD simulations were performed for a 1x2x1 supercell ("Small") and a 1x2x2 supercell ("Large"), both with periodic boundary conditions, at 520K using a Nosé-Hoover thermostat. The initial structure for each run was obtained by relaxing and statically optimizing the defect-free structure prior to removing a $Li^+$ ion. Γ-point meshes of 3x3x3 and 2x2x2 were used respectively for the "Small" and "Large" simulations. The plane-wave cutoffs for both systems were set to 450 eV. In both runs, the temperature of the system was initialized at 100K and then heated up to the target temperature of 520K at a constant rate by rescaling the velocity over a time span of 4 ps. Each system was then equilibrated at the target temperature for another 4 ps. After equilibrating, the "Small" trajectory was simulated for approximately 50 ps while the "Large" trajectory was simulated for approximately 14 ps, both using a timestep of 2 fs. Data used for training and testing the GNNFF only includes the trajectory generated after equilibration. In both trajectories, no breaking of the P-S bond was observed throughout the simulation.

**Data Availability**

The data that support the findings of this study are available from the corresponding author upon reasonable request. The training data for $Li_4P_2O_7$ and $Al_2O_3$-HF solid-state system is available through a Code Ocean compute capsule (https://doi.org/10.24433/CO.2788051.v1).

**Acknowledgements**

This work was performed in and funded by Bosch Research and Technology Center. This work was partially supported by ARPA-E Award No. DE-AR0000775. This research used resources of the Oak Ridge Leadership Computing Facility at Oak Ridge National Laboratory, which is supported by the Office of Science of the Department of Energy under Contract DE-AC05-00OR22725. CWP and CW also acknowledge financial assistance from Award No.70NANB14H012 from US Department of Commerce, National Institute of Standards and Technology as part of the Center for Hierarchical Materials Design (CHiMaD) and the Toyota Research Institute (TRI). The authors also thank Eric Isaacs and Yizhou Zhu for helpful discussion.

**Competing Interests**

The authors declare no competing interests

**Author Contributions**

JPM conceived the project and jointly developed the method with CWP. CWP implemented the GNNFF model with help from JPM and MK. MK, CWP, and JPM generated the data that were used for evaluations. JV, JPM, and CWP built the GNNFF MD engine. CWP performed the GNNFF MD study and error analysis with help from JPM, MK and BK. BK mentors the research at Bosch and is the primary academic advisor of JV. CW is the primary academic advisor for CWP. CWP wrote the manuscript. All authors contributed to manuscript preparation.

# Supplementary Information: Accurate and scalable multi-element graph neural network force field and molecular dynamics with direct force architecture

Cheol Woo Park[1,2], Mordechai Kornbluth[1], Jonathan Vandermause[3], Chris Wolverton[2], Boris Kozinsky[3,1*], Jonathan P. Mailoa[1*]

1) Robert Bosch Research and Technology Center, Cambridge, MA 02139, USA
2) Northwestern University, Evanston, IL 60208, USA
3) Harvard School of Engineering and Applied Sciences, Cambridge, MA 02138, USA

* corresponding authors: jpmailoa@alum.mit.edu, bkoz@seas.harvard.edu

**Supplementary Table S1.**
Here, we summarize the mean absolute value (MAV) and the GNNFF vector mean absolute error (vMAE) for each element-type of all solid-state systems presented in the manuscript. With the exception of $Al_2O_3$-HF (Reaction), for each system listed in the table, GNNFF was trained on 80% of the snapshots randomly chosen from the MD trajectory and tested on the other 20% of the snapshots. For $Al_2O_3$-HF (Reaction), GNNFF was trained on snapshots that do not correspond to the HF acid reacting with $Al_2O_3$ and tested on the snapshots that do correspond to the reaction. For each system, MAV and vMAE values were measured from the test set.

| System | Atom | MAV (eV/Å) | vMAE (eV/Å) | vMAE/MAV (%) |
|---|---|---|---|---|
| $Li_4P_2O_7$ | Li | 1.616 | 0.206 | 13 |
| | O | 3.440 | 0.356 | 10 |
| | P | 5.875 | 0.599 | 10 |
| $Al_2O_3$-HF (Standard) | Al | 1.999 | 0.290 | 14 |
| | F | 1.698 | 0.235 | 14 |
| | H | 1.361 | 0.188 | 14 |
| | O | 1.756 | 0.260 | 15 |
| $Al_2O_3$-HF (Reaction) | Al | 2.093 | 0.356 | 17 |
| | F | 1.604 | 0.297 | 19 |
| | H | 1.466 | 0.235 | 16 |
| | O | 1.788 | 0.299 | 17 |
| $Li_{7-x}P_3S_{11}$ (Small) | Li | 0.487 | 0.087 | 18 |
| | P | 1.540 | 0.194 | 13 |
| | S | 0.958 | 0.172 | 18 |
| $Li_{7-x}P_3S_{11}$ (Large) | Li | 0.467 | 0.097 | 21 |
| | P | 1.553 | 0.244 | 16 |
| | S | 0.951 | 0.199 | 21 |